\documentclass[11pt]{article}
\usepackage{moriond,epsfig}
\usepackage{units}

\begin{document}
\vspace*{4cm}
\title{Investigation of Hadronic Interactions at Ultra-High Energies with the\\ Pierre Auger Observatory}

\author{R. Ulrich$^{1}$ for the Pierre Auger Collaboration$^{2}$}

\address{\vspace*{.3cm} $^{1}$Karlsruhe Institute of
  Technology~\footnote{KIT is the cooperation of University Karlsruhe
    and Forschungszentrum Karlsruhe} (KIT) \\ Institut f\"ur
  Kernphysik, P.O. Box 3640, 76021 Karlsruhe,
  Germany\\[.2cm]$^{2}$Observatorio Pierre Auger, Av. San Martin Norte
  304, Malarg\"ue, Argentina}

\maketitle\abstracts{ The aim of the Pierre Auger Observatory is the
  investigation of the nature of cosmic ray particles at ultra-high
  energies. It can simultaneously observe the longitudinal air shower
  development in the atmosphere as well as particle densities on the
  ground. While there are no dedicated muon detectors, techniques have
  been developed to estimate the number of muons, $N_\mu$, produced by
  air showers.  Both, the longitudinal development, in particular the
  depth of the shower maximum, $X_{\rm max}$, and the muon content of
  air showers are highly sensitive to hadronic interactions at
  ultra-high energies. Currently, none of the available hadronic
  interaction models used for simulations of extensive air showers is
  able to consistently describe the observations of $X_{\rm max}$ and
  $N_\mu$ made by the Pierre Auger Observatory.}

%
\section{Introduction}
At the Pierre Auger Observatory~\cite{Abraham:2004dt} extensive air
showers induced by ultra-high energy cosmic ray particles with
energies of $10^{18}-10^{20}\,$eV are investigated. The aim of the
experiment is to solve the mysteries on the nature and the sources of
these ultra-high energy cosmic ray particles. Since individual cosmic
ray primaries and also ultra-high energy secondaries in the startup of
the air shower cascade are interacting with our atmosphere at
center-of-mass energies up to $\unit[\sim450]\,$TeV, the Pierre Auger
Observatory is also sensitive to the physics of hadronic interactions
at center-of-mass energies far beyond the reach of the LHC. We
demonstrate the current ability to test existing hadronic interaction
models at these energies.

\begin{figure}[bt!]
  \centering
  \includegraphics[width=.49\linewidth]{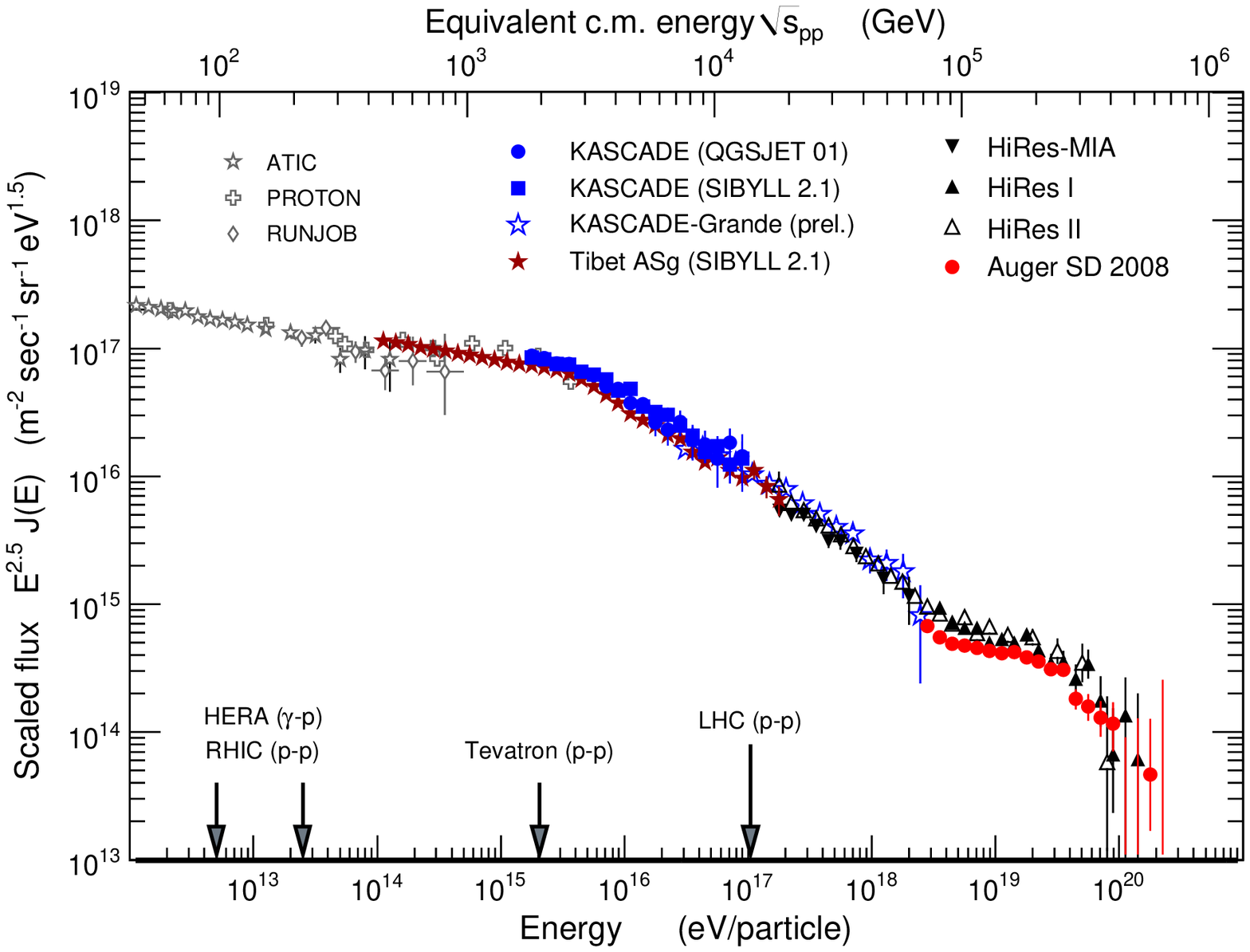}~
  \includegraphics[width=.51\linewidth]{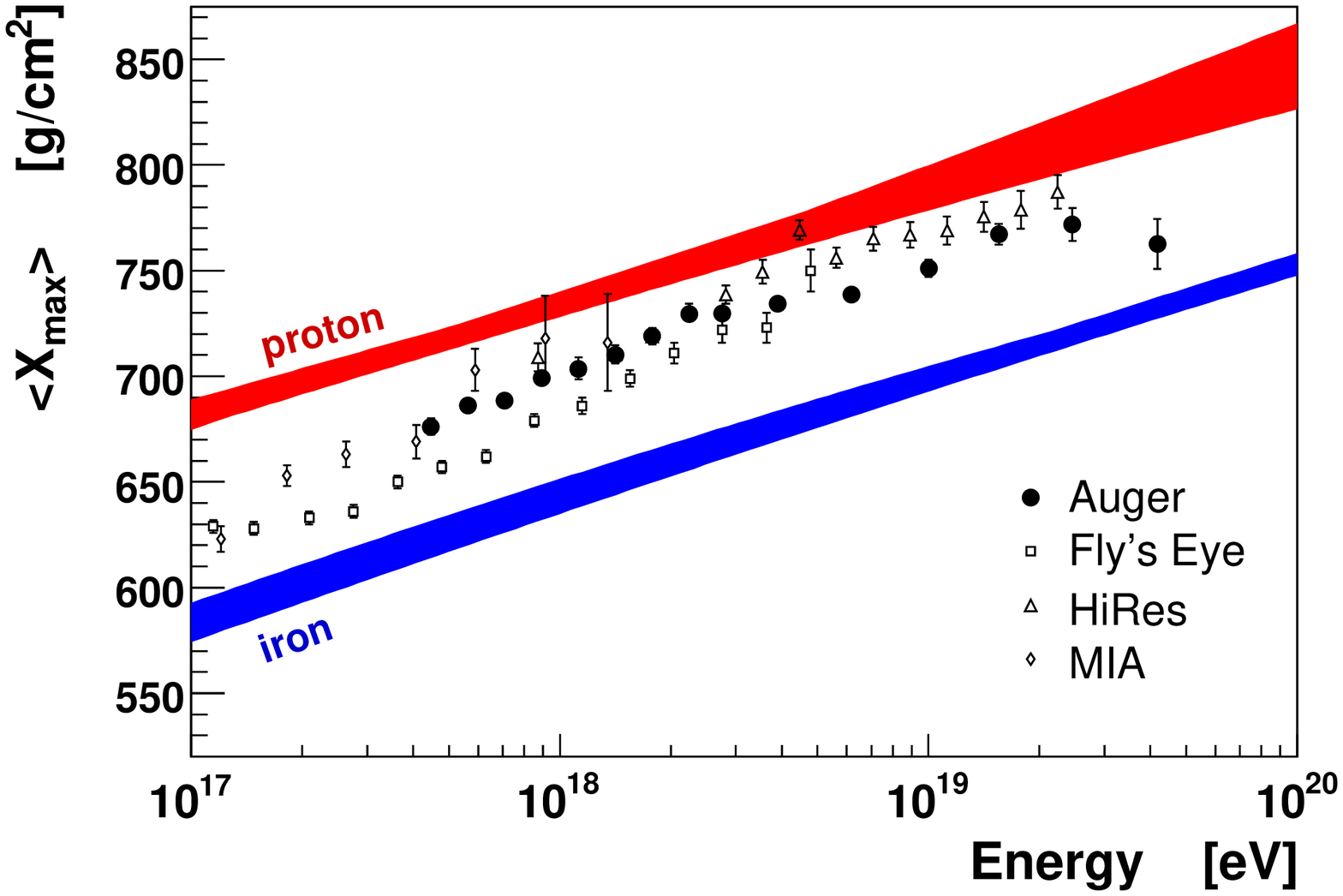}\\
  \vspace*{-.3cm}
  \caption{Left panel: The total flux of cosmic ray
    particles~\protect\cite{Abraham:2008ru,Bluemer:2009zf} multiplied
    by the energy to the power of 2.5.  Right panel: Measurements of
    $\langle X_{\rm max}\rangle$ compared to
    simulations~\protect\cite{AugerXmax}. The uncertainty bands for
    proton and iron are indicating the spread caused by the hadronic
    interaction models
    \textsc{QGSJet01}~\protect\cite{Kalmykov:1989br}, 
    \textsc{QGSJetII}~\protect\cite{Ostapchenko:2004ss}, 
    \textsc{Sibyll}~\protect\cite{Fletcher:1994bd} and 
    \textsc{Epos}~\protect\cite{Werner:2007vd}.}
  \vspace*{-.1cm}
  \label{fig:flux}
\end{figure}

The observatory consists of an array of 1600 water-Cherenkov particle
detectors distributed on a surface of \unit[3000]{km$^2$} 
combined with 24 fluorescence telescopes overlooking the
atmosphere above the surface array in moonless nights. While the
advantage of the surface array is a duty cycle of close to 100\,\%, the
fluorescence telescopes, which have a duty cycle of
\unit[$\sim$13]{\%}, provide an almost calorimetric measurement of the
energy of the air showers as well as the position of the maximum energy
deposition in the atmosphere, $X_{\rm max}$.

The Pierre Auger Observatory combines the strengths of the two
detection techniques, and can thus achieve an unpreceeded level of
quality in data analysis. A good example is the measurement of the
total flux, which is founded on the purely geometric acceptance of the
surface array combined with the almost calorimetric energy
reconstruction of the fluorescence
telescopes~\cite{Abraham:2008ru}. This makes it the first cosmic ray
experiment that is able to reconstruct the cosmic ray flux with an
absolute minimum of systematic uncertainties induced by Monte
Carlo. The resulting total cosmic ray flux is displayed in
Figure~\ref{fig:flux}~(left), where it is shown together with the
results of other experiments.  The spectrum in the energy region of
the Pierre Auger Observatory exhibits two important features. Around
$\unit[10^{18.5}]{eV}$ there happens a hardening of the spectrum, and
above $\approx\unit[10^{19.6}]{eV}$ a suppression is observed. The
former is believed by most theorists to be related to the transition
from a galactic to an extragalactic origin of cosmic
rays~\cite{Kulikov1959,Hillas:2005cs,Allard:2007gx,Berezinsky:2005cq}. The
latter is compatible with the GZK
cutoff~\cite{Zatsepin:1966jv,Greisen:1966jv}, which is caused by
interactions of cosmic ray particles with the extragalactic photon
field, while it could also be related to the maximum acceleration
energy of the cosmic ray source(s).

On the other hand, for other analyses, as for example the
reconstruction of the primary mass composition of cosmic rays, the
prediction of air shower Monte Carlo simulations are a crucial
ingredient~\cite{Knapp:2002vs}. The poorly constraint physics of
hadronic interactions at the relevant ultra-high energies and phase
space regions are currently preventing an unambiguous, or even
self-consistent, analysis of the available data.
At the same time it is demonstrated that air shower observables are
very sensitive to hadronic interaction
physics~\cite{Ulrich:2009zzz}. Thus, by using astrophysical
constraints on the primary cosmic ray mass composition it is possible
to study the features of hadronic interactions at ultra-high energies
occuring in the startup of extensive air showers. The observables most
sensitive to hadronic interaction physics are the depth of the shower
maximum, $X_{\rm max}$, and the number of muons on the observation
level, $N_\mu$.

%
\section{The depth of the shower maximum}
The depth of the shower maximum, $X_{\rm max}$, is one of the most
important observables of the telescope detectors. It is reconstructed
by fitting a Gaisser-Hillas parameterization to the reconstructed
energy deposit profiles~\cite{Unger:2008uq}. With the Pierre Auger
Observatory a resolution better than $\unit[20]{g/cm^2}$ for $X_{\rm
  max}$ is achieved~\cite{Dawson:2007di}. In
Fig.~\ref{fig:flux}~(right) the measured mean $X_{\rm max}$ versus
energy is compared to predictions from air shower Monte Carlo
simulations for different primary cosmic ray particles and hadronic
interaction models. The main features of the data of the Pierre Auger
Observatory are a slight trend from heavier to lighter composition at
low energies, while above $10^{18.2}\,$eV this trend is stopped, or
maybe even reversed. The data can be described by all hadronic
interaction models with a medium or mixed mass composition with
$A\sim7$.

%
\section{The number of muons at the observation level}
\begin{figure}[bt!]
  \centering
  \includegraphics[width=.5\linewidth]{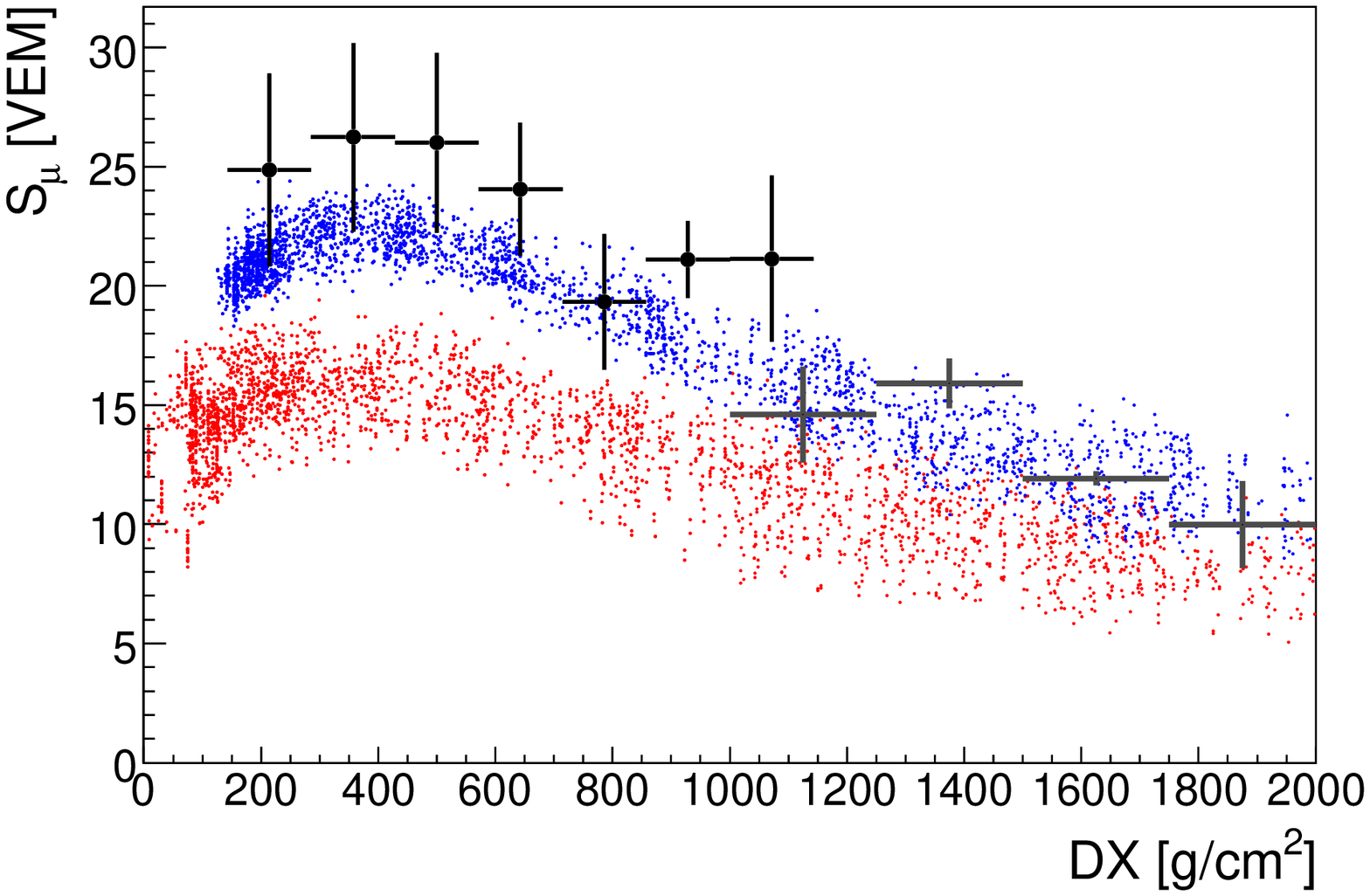}~
  \includegraphics[width=.5\linewidth]{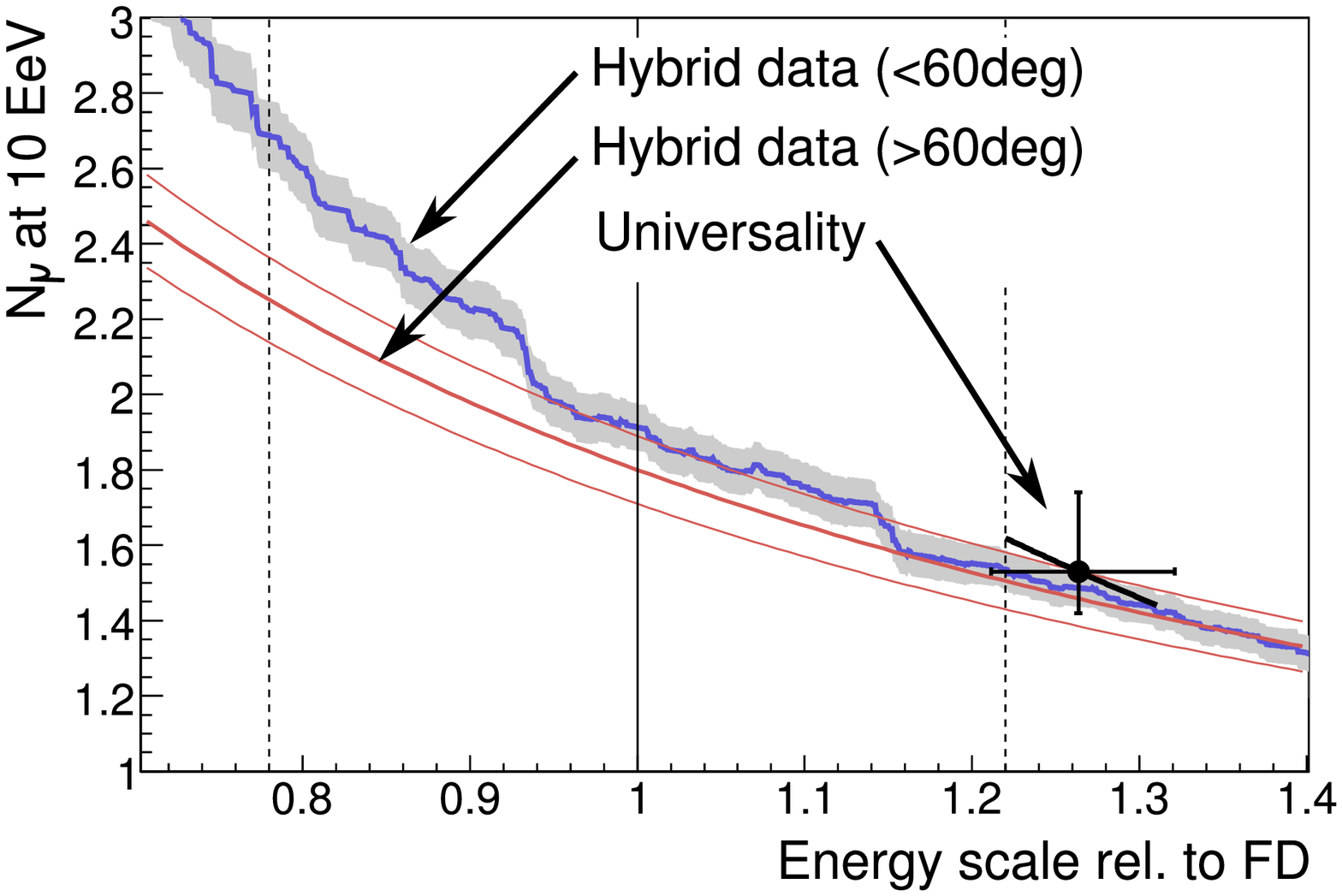}\\
  \vspace*{-.3cm}
  \caption{Left panel: Measured muonic signal, $S_\mu$, in
    water-Cherenkov tanks obtained after subtraction of the
    electromagnetic signal component~\protect\cite{Engel:2007cm}. The
    lower markers are the predictions by \textsc{QGSJetII} for proton
    and the upper markers for iron induced air showers. The energy of
    the simulated showers was shifted up by 30\,\%. Right panel:
    Comparison of the muon reconstruction methods applied to Auger
    data. The muon number $N_\mu$ is defined as the relative muon
    number compared to the predictions of \textsc{QGSJetII}, and the
    energy scale may be varied within the systematic uncertainties,
    which are indicated as dashed vertical
    lines~\protect\cite{Engel:2007cm,Schmidt:2007vq,Schmidt:2009ge}.}
  \vspace*{-.1cm}
  \label{fig:muon1}
\end{figure}
The Pierre Auger Observatory has no dedicated muon detectors. However,
it is possible to indirectly infer the number of muons in the
water-Cherenkov tanks of the surface array with several
methods~\cite{Engel:2007cm,Schmidt:2007vq,Schmidt:2009ge}. The
principle of one technique, which combines the data of the telescopes
with that of the surface array, is the subtraction of the
electromagnetic part of the air shower, as it is measured by the
telescopes, from the signal in the surface array to yield the muon
signal. The results of this method are shown in
Fig.~\ref{fig:muon1}~(left).  Even with the energy of the simulations
scaled up by 30\,\%, which is just within the systematic uncertainties
quoted by the Auger Collaboration~\cite{Dawson:2007di}, the data are
lying beyond the iron predictions. Similar results are obtained with
all the other muon measurement methods on the data of the Pierre Auger
Observatory, which are summarized in Fig.~\ref{fig:muon1}~(right).
The muon number larger than the predictions by iron primaries is by
itself a strong indication of a deficient hadronic interaction
modeling. Furthermore, comparing the results to the $\langle X_{\rm
  max}\rangle$ data exhibits a striking incompatibility of the
interpretations with the existing interaction models.

All the results discussed here are based on comparisons to simulations
performed with \textsc{QGSJetII}. The choice of a different
interaction model does not change the findings qualitatively. Even the
\textsc{Epos} model, which predicts the largest muon
numbers~\cite{Pierog:2006qv}, cannot fully account for the
discrepancies.

%
\section{Summary}

It is shown that the two observables of the Pierre Auger Observatory
that are most sensitive to hadronic interaction physics at ultra-high
energies, $X_{\rm max}$ and $N_\mu$, do not give a consistent
picture when compared to air shower Monte Carlo simulations. While
the $\langle X_{\rm max}\rangle$ data is located well in between the
predictions for proton and iron, the $N_\mu$ data is located 
outside of this interval and would favor a mass composition even 
heavier than iron.

This result clearly points out the deficiency of existing hadronic
interaction models to describe cosmic ray data at ultra-high
energies. The characteristics of hadronic interactions within the air
shower cascade must be different from what is currently assumed in all
the models.

With improved constraints on the primary cosmic ray mass composition,
observations of this kind can be used to study the relevant properties
of hadronic interaction physics at center-of-mass energies up to
$\sim450\,$TeV.

%
\section*{Acknowledgments}
I thank the members of the Pierre Auger Collaboration for discussions
and a general very productive and stimulating scientific environment. 

%
\section*{References}
--------------------------------------------------------------------------
\bibliographystyle{unsrt-mod-notitle}
\bibliography{moriond}
%

\end{document}